\documentclass[preprint,12pt]{elsarticle}

\usepackage{amssymb}
\usepackage[utf8]{inputenc}
\usepackage{soul}
\usepackage{color}
\usepackage{graphicx}
\usepackage{amsmath}
\usepackage[version=4]{mhchem}
\usepackage{siunitx}
\usepackage{longtable,tabularx}
\usepackage{caption}
\usepackage{float}
\usepackage[nameinlink,capitalise,noabbrev]{cleveref}
\usepackage[dvipsnames]{xcolor}
\usepackage{algorithm}
\usepackage{algorithmic}
\graphicspath{{Figures/}}
\usepackage{subfigure}
\usepackage{subcaption}
\usepackage{multirow}
\usepackage{booktabs}
\usepackage{xcolor}

\begin{document}

\begin{frontmatter}

\title{Multi--Fidelity Bayesian Neural Network for Uncertainty Quantification in Transonic Aerodynamic Loads}

\author[inst1]{Andrea Vaiuso\corref{cor1}\fnref{label2}}
\fntext[label2]{Research Associate}
\ead{Andrea.Vaiuso@zhaw.ch}

\cortext[cor1]{Corresponding Author}

\author[inst1,inst2]{Gabriele Immordino\fnref{label2}}
\fntext[label1]{Ph.D. Student}

\author[inst1]{Marcello Righi\fnref{label4}}
\fntext[label4]{Professor, AIAA Member, Lecturer at Federal Institute of Technology Zurich ETHZ}

\author[inst2]{Andrea Da Ronch\fnref{label3}}
\fntext[label3]{Associate Professor, AIAA Senior Member}

\affiliation[inst1]{organization={School of Engineering, Zurich University of Applied Sciences ZHAW},
            city={Winterthur},
            country={Switzerland}}

\affiliation[inst2]{organization={Faculty of Engineering and Physical Sciences, University of Southampton},
            city={Southampton},
            country={United Kingdom}}

\begin{abstract}

Multi-fidelity models are becoming more prevalent in engineering, particularly in aerospace, as they combine both the computational efficiency of low-fidelity models with the high accuracy of higher-fidelity simulations. Various state-of-the-art techniques exist for fusing data from different fidelity sources, including Co-Kriging and transfer learning in neural networks. This paper aims to implement a multi-fidelity Bayesian neural network model that applies transfer learning to fuse data generated by models at different fidelities. Bayesian neural networks use probability distributions over network weights, enabling them to provide predictions along with estimates of their confidence. This approach harnesses the predictive and data fusion capabilities of neural networks while also quantifying uncertainty. The results demonstrate that the multi-fidelity Bayesian model outperforms the state-of-the-art Co-Kriging in terms of overall accuracy and robustness on unseen data.


\end{abstract}

\end{frontmatter}

\section{Introduction}

In recent years, there has been increasing interest in aerospace engineering in combining data from simulation models of varying fidelities. This approach aims to improve the accuracy and efficiency of predicting physical behaviors, including aerodynamic loads, aeroelasticity, flight dynamics, and so forth \cite{yamazaki2013derivative, fernandez2016review, brevault2020overview, wilke2024multifidelity}. 

In general, low-fidelity models are employed to calculate physical quantities of complex systems in a simplified manner in situations where rapid assessments are required~\cite{crovato2020effect,carrizales2020verification,taylor2021low}. This is achieved through the identification of the key physical phenomena, the application of simplifying assumptions to reduce degrees of freedom, and the use of basic mathematical models or empirical data to approximate the behaviour of the system~\cite{lucia2004reduced, kim2005aeroelastic,wang2008flutter,morino2009efficient,mannarino2015reduced}. These models, such as simplified vortex models~\cite{segalini2013simplified,xu2018simplified} and reduced-order models~\cite{lucia2004reduced,lassila2014model,levin2022convolution,dowell2023reduced}, provide essential aerodynamic characteristics without incurring the significant computational demands of high-fidelity simulations like Computational Fluid Dynamics (CFD), which accurately model complex flow behaviors including turbulence and viscous effects~\cite{rodriguez2019applied}. However, these simplified models often struggle to achieve sufficient accuracy when predicting regions of the flight envelope that exhibit highly nonlinear behavior~\cite{eivazi2020deep, zhang2023review}.

To bridge this gap, a valid approach is represented by Data Fusion (DF) techniques, which aim to leverage the strengths of models based on different fidelities to create a multi-fidelity model~\cite{he2020multi,romor2023multi}. A common DF approach is to generate this kind of model by combining a data-driven model based on low-fidelity samples with a small number of high-fidelity points, with the goal of improving the overall prediction accuracy. This approach is particularly beneficial when the generation of high-fidelity data about a specific physical phenomenon is computationally expensive, yet low-fidelity and inexpensive models are available. A state-of-the-art DF technique, Co-Kriging, utilizes Gaussian Process (GP) to correct low-fidelity predictions using correlations between different fidelity levels, thereby improving accuracy and quantifying uncertainty effectively~\cite{liu2018sequential,pham2023extended}. Forrester et al.\cite{forrester2007multi} demonstrated Co-Kriging capability in multi-fidelity wing optimization, highlighting its utility in integrating diverse data sources for aerodynamic studies, a method still prevalent in recent aerodynamic studies and airfoil optimizations~\cite{da2011generation, han2012hierarchical, liu2018sequential}. 



The integration of machine learning (ML) has significantly advanced the DF field, offering rapid and accurate predictions despite the initial training challenges, particularly with Deep Learning (DL)~\cite{sharma2022machine,stahlschmidt2022multimodal}. The demand for large quantities of high-quality training data, often limited and costly to obtain through CFD, underscores the importance of DF techniques for leveraging multi-fidelity sources. Transfer Learning (TL) is a frequently employed approach in DL networks for DF tasks, involving initial training with low-fidelity data followed by fine-tuning with sparse high-fidelity samples to enhance accuracy~\cite{zhuang2020comprehensive,li2022line}. In \cite{chakraborty2021transfer}, a multi-fidelity physics-informed deep neural network uses TL to predict reliability analysis outcomes effectively, surpassing standalone models, while in \cite{liao2021multi}, TL integrates low-fidelity and high-fidelity data to improve a CNN-based model accuracy for aerodynamic optimization.

ML-based approaches have shown superior performance over traditional methods like Co-Kriging in capturing data non-linearity \cite{eivazi2020deep, guo2022multi, rakotonirina2024generative}. However, many ML-based DF models currently lack direct mechanisms for uncertainty quantification, both in the model and the underlying data, posing a significant challenge for practical applications. Bayesian Neural Networks (BNNs) address this limitation by incorporating Bayesian inference to estimate distributions over network weights~\cite{tran2019bayesian}, providing probabilistic interpretations of predictions that capture both model and data uncertainties, making them particularly suitable for applications in aerospace engineering where understanding the reliability of predictions is crucial~\cite{huang2020bayesian}. Recent research, such as Meng et al.~\cite{meng2021multi} and Sharma et al.~\cite{sharma2022exploring}, explores BNNs in multi-fidelity models, highlighting challenges in computational complexity and data integration mismatches. Kerleguer et al.~\cite{kerleguer2024bayesian} propose a hybrid approach combining GP regression and BNNs to mitigate these challenges, demonstrating promising avenues for integrating diverse data sources effectively.

The aim of our paper is to develop a ML-based framework for multi-fidelity using exclusively BNN layers, and integrating TL for the data fusion process. This approach has led to the design of a more straightforward architecture, solely based on the DL paradigm. This framework will be implemented using open-source code to ensure replicability and accessibility for further research and applications. 

This paper is structured as follow: Section~\ref{sec:meth} introduces the concepts of BNNs and TL, and presents the Multi-Fidelity Bayesian Neural Network with Transfer Learning (MF-BayNet) model. Section~\ref{sec:studycase} outlines the Benchmark Super Critical Wing (BSCW) case study and the methodology adopted for generating the transonic aerodynamic loads with different fidelities. Section~\ref{sec:results} discusses the results obtained with the framework trained using numerous low-fidelity samples and few high-fidelity CFD points, demonstrating the superior performance of the MF-BayNet model over models trained on each dataset separately, as well as compared to Co-Kriging, while also providing reliable uncertainty quantification results. Finally, Section~\ref{sec:concl} summarizes the conclusions drawn from this study and outlines directions for future research.


\section{Methodology}\label{sec:meth}

This section details the steps involved in creating the model. We implemented a multi-fidelity framework integrating Bayesian neural networks with transfer learning techniques to harness diverse data sources and enhance model generalization. The approach includes rigorous quantification of uncertainty to ensure reliable predictions, along with systematic optimization of model hyperparameters for optimal performance. Furthermore, Co-kriging model is introduced in order to perform a comparative analysis, benchmarking the efficacy of our proposed method.

\subsection{Bayesian Neural Networks}
Bayesian Neural Networks (BNNs) extend standard neural networks by integrating Bayesian inference, which allows for the estimation of uncertainty in the model predictions. Unlike traditional neural networks that only provide sample estimates, BNNs generate a distribution over the network weights, capturing both model and data uncertainty. 
In practical, the Bayesian approach involves specifying a prior distribution $p(\theta)$ over the network weights $\theta$. The training phase is intended to update this distributions with the observed data 
$\mathcal{D}$ to obtain a posterior distribution 
$p(\theta | \mathcal{D})$. 

For regression tasks, the BNN outputs a distribution over predictions. The predictive distribution is obtained by marginalizing over the posterior distribution of the weights, obtaining:
\begin{equation}
    p(y | x, \mathcal{D}) = \int p(y | x, \theta) p(\theta | \mathcal{D}) d\theta
\end{equation}

Learning this model is often intractable because computing the posterior involves integrating over the entire parameter space, a task that typically lacks a closed-form solution. There are several methods for approximating the posterior, including the Laplace Method \cite{10.1162/neco.1992.4.3.448}, which approximates the posterior with a Gaussian distribution; Hamiltonian Monte Carlo (HMC) \cite{neal1992bayesian}, which uses Hamiltonian dynamics for sampling; and others techniques based on Markov Chain Monte Carlo \cite{chen2014stochastic}.

The posterior distribution can be efficiently approximated using variational inference \cite{graves2011practical}, which minimizes the Kullback-Leibler (KL) divergence between the true value and the posterior, and an approximate distribution $q(\theta)$. KL divergence between the two distributions is defined as:
\begin{equation}
    \text{KL}(q(\theta) \| p(\theta | \mathcal{D})) = \int q(\theta) \log \frac{q(\theta)}{p(\theta | \mathcal{D})} d\theta
\end{equation}
In BNNs that uses variational inference for solving the posterior, the loss function is defined using KL divergence, combining the mean squared error (MSE) for the regression task with the KL divergence term. The overall loss function can be expressed as:
\begin{equation}
    \mathcal{L}(\theta) = \mathcal{L}_{\text{MSE}} + \text{KL}(q(\theta) \|p(\theta | \mathcal{D}))
\end{equation}
Where $\mathcal{L}_{\text{MSE}}$ is the mean squared error loss, and $\text{KL}(q(\theta) \| p(\theta | \mathcal{D}))$ is the KL divergence between the approximate posterior and the prior distribution.

In our framework, we built the BNN model using Pytorch and Torchbnn library, that provides a fully customizable Pytorch framework for building a BNN architectures based on variational inference.

\subsection{Multi-Fidelity BNN-TL}

The core of our methodology revolves around a multi-fidelity model that combines BNNs with TL, referred as MF-BayNet in the rest of the paper. The process begins by training a BNN on a large dataset of low-fidelity data to capture general phenomena trends. Subsequently, the model undergoes TL on mid-fidelity data, a common ML technique where a model developed for one task is reused as the starting point for a model addressing a second task. This approach leverages the knowledge gained from solving one problem to tackle related problems. Finally, the model is fine-tuned using few high-fidelity data points.

\subsubsection*{Transfer Learning Process}

In the context of data fusion, once the model has captured the basic understanding of the general behaviors during initial training, the final layers of the model are frozen by increasing the fidelity of the training set. This means that all frozen neurons have fixed values of weights during training, which in the case of BNN are represented by the mean and standard deviation of each probability distribution. Then, the subset of trainable parameters of the model is retrained on one or more small sets of increasingly higher fidelity data. A schematic of the architecture is illustrated in Figure \ref{fig
}.

\begin{figure}[!ht]
\centering
\includegraphics[width=1\linewidth]{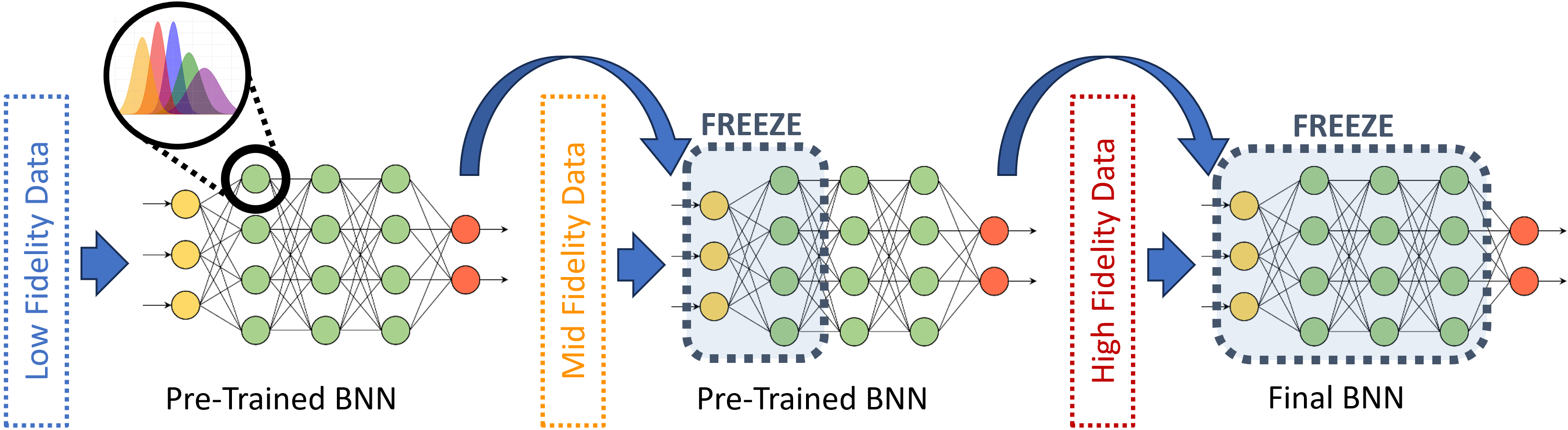}
\caption{Architecture schema of MF-BayNet}
\label{fig
}
\end{figure}

The TL approach significantly reduces the computational cost and time required for training thanks to the reduced subset of trainable neurons, while improving the model ability to make accurate predictions using different fidelities and emphasizing the importance of higher fidelity data. In addition, this method relies only on DL principles, which makes it more straightforward and robust compared to other methods. Two TL processes were executed: the first on mid-fidelity data and the second on high-fidelity data points to fine-tune the model. In the second process, fewer layers were frozen compared to the first, allowing for more refined adjustments.

\subsubsection*{Prediction Means and Confidence Interval}

Once the TL process is completed, we use Monte Carlo Sampling~\cite{caflisch1998monte} to obtain the prediction means and standard deviations from the model, which are used for estimating the model uncertainty and reliability in its predictions. First, multiple forward passes are performed through the BNN. Each pass generates a different set of weights due to the nondeterministic behavior of the model, effectively creating an ensemble of models. Next, the outputs of these forward passes are averaged to obtain the mean prediction. This gives us an estimate of the expected value of the prediction. Finally, the standard deviation of the outputs from the multiple forward passes is calculated to assess the uncertainty in the predictions. This standard deviation provides a measure of how much the predictions vary, offering insights into the confidence of the model. This value represents the predictive uncertainty, a term that contains both epistemic and aleatory uncertainty quantification.

\subsubsection*{Uncertainty Quantification}

Uncertainty quantification is crucial for understanding the reliability of the model predictions. The MF-BayNet model provides a probabilistic interpretation of predictions, offering insights into both model and data uncertainty. The total uncertainty in the prediction, the predictive uncertainty ($P_u$), is defined as the sum of
epistemic ($P_u$) and aleatory uncertainty ($A_u$)~\cite{chai2018uncertainty, depeweg2017uncertainty, der2009aleatory}.

\begin{equation}\label{eq:uq}
    P_u=E_u+A_u
\end{equation}

Epistemic uncertainty refers to the uncertainty in the model parameters. This can be visualized as the spread of the posterior weight distribution $p(w|D)$. In ML, this type of uncertainty emerges when the model has not encountered data that adequately represents the entire design domain, or when the domain itself needs further refinement or completion. This type of uncertainty arises due to deficiencies from a lack of knowledge or information~\cite{roadmap2024easa}. In contrast, aleatory uncertainty arises from the inherent variability in the input data. Given a specific input and fixed weight parameters, high aleatory uncertainty indicates that the output estimate is subject to noise. This kind of uncertainty refers to the intrinsic randomness in the data, which can derive from factors such as data collection errors, sensor noise, or noisy labels~\cite{roadmap2024easa}. High aleatory uncertainty suggests insufficient information to predict the output accurately due to unobserved or latent variables that the model cannot capture. Predictive uncertainty alone cannot distinguish between different types of uncertainty. Inputs with high predictive uncertainty may have contributions from epistemic uncertainty, aleatory uncertainty, or both.

In BNN, using different weight settings could produce very confident,
but disagreeing predictions. In addition, after running Monte Carlo on the weights, it is not possible to directly distinguish between the uncertainty associated with the weights and the intrinsic uncertainty of the input. Furthermore, the TL process leads to a change in the aleatory uncertainty dependence of different fidelity inputs. In fact, only the subset of neurons with trainable parameters holds the probability information about inputs of different fidelities. The possible high discrepancy between datasets of different fidelities suddenly increases the aleatory uncertainty of the process, but a higher fidelity training dataset should, by definition, contain more reliable samples compared to lower fidelity ones. 
It is important to consider that it is frequently not feasible to ascertain the reliability of a particular simulation model a priori. Consequently, it is not possible to assess the impact of a single alteration to the uncertainty during the TL process. 


\subsection{Model Optimization}

Choosing the right hyperparameters of the network is a complex task. This requires a deep understanding of the model architecture and the specific characteristics of the data at different fidelity levels in order to create a good hyperparameter optimization process. We implemented a Bayesian Optimization~\cite{snoek2012practical} strategy to obtain the best hyperparameters set for each model. Bayesian optimization strength lies in its iterative approach to fine-tuning hyperparameters using Bayesian probability distributions, rather than exhaustively testing every possible combination. Each iteration, known as a trial, involves training the network with a defined set of hyperparameters and optimizing them based on past trials performance with respect to the validation set metric. This cycle repeats until the optimal outcome is attained. The pseudo-code of the Bayesian optimization strategy is depicted in Algorithm~\ref{alg:bayesian_optimization_kt}.

\begin{algorithm}[!b]
\caption{Bayesian Optimization for Hyperparameter Tuning}
\label{alg:bayesian_optimization_kt}
\begin{algorithmic}[1]
\STATE \textbf{Input:} Objective function $f$, search space $S$, hyperparameter tuner $T$, number of trials $N_{\text{trials}}$
\STATE \textbf{Output:} Optimal hyperparameter set $\theta^*$
\STATE Initialize hyperparameter tuner $T$ with search space $S$
\STATE Perform an initial random search to populate $T$

\STATE Initialize optimal hyperparameter set: $\theta^* \leftarrow \text{None}$
\STATE Initialize optimal objective value: $y^* \leftarrow -\infty$

\FOR{$i = 1$ to $N_{\text{trials}}$}
    \STATE Sample the next hyperparameter set from $T$: $\theta_i \leftarrow \text{sample}(S)$ 
    \STATE Evaluate the objective function: $y_i = f(\theta_i)$
    
    \IF{$y_i > y^*$}
        \STATE Update the optimal hyperparameter set: $\theta^* \leftarrow \theta_i$
        \STATE Update the optimal objective value: $y^* \leftarrow y_i$
    \ENDIF
    
    \STATE Update the posterior distribution based on observed data
    
    \STATE Select: $\theta_{\text{next}} \leftarrow \arg\max_{\theta} \text{AcquisitionFunction}(\theta; \text{Posterior})$
    
    \STATE Incorporate $\theta_{\text{next}}$ into $T$
\ENDFOR

\STATE \textbf{Return:} Optimal hyperparameter set $\theta^*$
\end{algorithmic}
\end{algorithm}

The design parameters targeted for the optimization process include the number of units per layer ($N_{units}$), the total number of layers in the model ($N_{layers}$), activation functions, optimization function, batch size, number of epochs, and the learning rate for each training phase. After the $i-th$ TL phase, the model needs to be retrained with a different learning rate value ($lr_i$). Other parameters include the prior distribution (initial probability distribution for each weight, determined by the mean ($\mu_{prior}$) and variance ($\sigma_{prior}$) values of a Gaussian function, and the number of layers to freeze during transfer learning ($N_{frz}$). The last one represents a critical task, as freezing too many layers might prevent the model from adapting to the new, higher-fidelity data, while freezing too few layers can lead to excessive retraining and potentially overfitting. The design space for all hyperparameters, including the possible values and step size for each variable, is presented in Table~\ref{tab_Hyperparameters_design_space}. A total number of 300 trials per model tested has been executed, with an average time per trial of 3 minutes.


\begin{table}[!ht]
\centering
\begin{tabular}{l c c }
\hline 
\hline 
\textbf{Hyperparameter} & \textbf{Value} & \textbf{Step size} \\ 
\hline
$N_{layers}$ & 3 to 6 & 1 \\
$N_{units}$ & 16 to 176 & 16 \\
$lr_i$ & $1\cdot 10^{-4}$ to $1\cdot 10^{-1}$ & $5\cdot 10^{-3}$   \\
$\mu_{prior}$ & -1.5 to 1.5  &  $5\cdot 10^{-2}$ \\
$\sigma_{prior}$ & $1\cdot 10^{-4}$ to $1\cdot 10^{-2}$  &  $5\cdot 10^{-4}$ \\
$N_{frz}$ & 1 to $(N_{layers}-1)$ & 1 \\
Activation $f$ & ReLU, PReLU, LeakyReLU  &  -- \\

\hline
\hline 
\end{tabular}
\caption{Hyperparameters design space.}
\label{tab_Hyperparameters_design_space}
\end{table}

During each trail of the optimization process, the model parameters, which are represented by the mean and variance values of the probability distributions of each neuron in the BNN model, along with bias values of the activation functions, are optimized based on the Mean Absolute Error (MAE) on the validation set.


\subsection{Co-Kriging}

Co-Kriging model is a state-of-the-art method for combining data from both low-fidelity and high-fidelity simulations, serving as a benchmark for comparison with our implemented methodology. The Co-Kriging function, denoted as $\hat{\eta}$, is computed from the samples of the low-fidelity evaluations, as outlined by Da Ronch et al.~\cite{da2011generation}. This function is then applied at the points where high-fidelity predictions are available. The resulting set of input parameters at the high-fidelity samples, represented as $x_i$, is expanded with the evaluation of the Co-Kriging function for the low-fidelity samples. The expanded vector has three dimensions, with $x_i^{aug}=[AoA_i, M_i, \hat{\eta}(x_i)]$. The corresponding aerodynamic loads for each of the $n_{i}$ sample points are represented as $y_i=y(x_i)$. Subsequently, a Co-Kriging function is computed for these expanded samples, $\hat{\eta}(x_i^{aug})$, with the additional component providing data for the correlation calculation from the low-fidelity samples. 

\section{Test Case}\label{sec:studycase}

\begin{figure}[!htb]
    \centering
\includegraphics[trim=0 0 0 0, clip, width=0.8\textwidth]{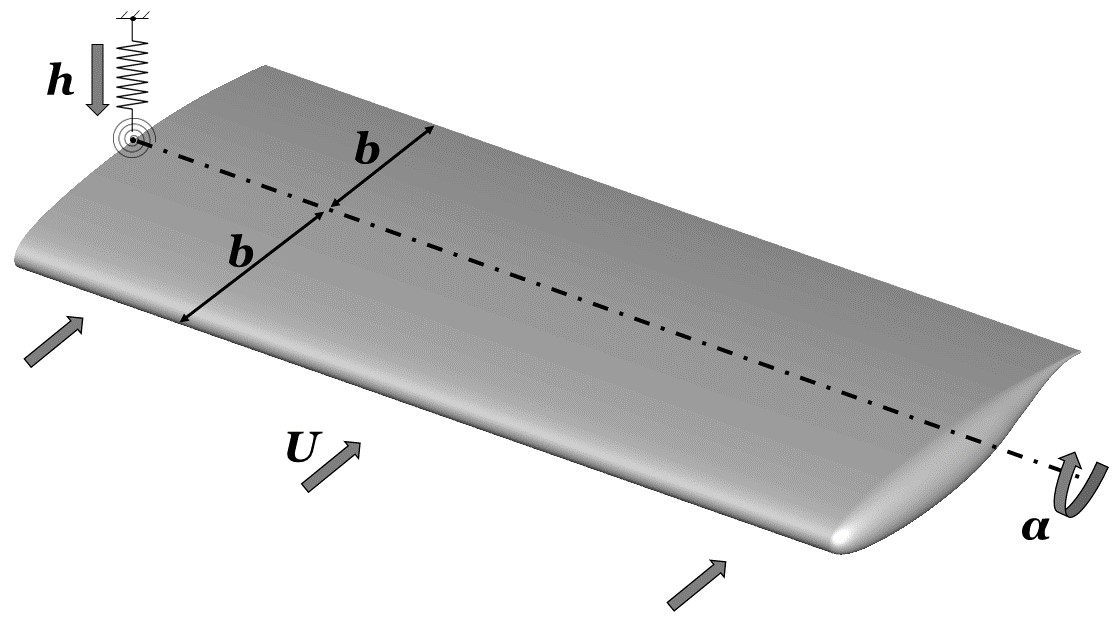}
\caption{ Schematic of the half-span BSCW.}
\label{fig:aeroelastic_bscw_sketch}
\end{figure}

This section details the practical application of the MF-BayNet model to predict aerodynamic loads on the Benchmark Super Critical Wing (BSCW) using a combination of low-fidelity, mid- and high-fidelity datasets. The BSCW, featured in the AIAA Aeroelastic Prediction Workshop \cite{heeg2013overview}, is a transonic, rigid, semi-span wing with a rectangular planform and a supercritical airfoil shape which is elastically suspended on a flexible mount system with two degrees of freedom, pitch and plunge. It exhibits complex aerodynamic phenomena such as shock wave motion, shock-induced boundary-layer separation, and interactions between shock waves and detached boundary layers. These nonlinearities present significant challenges for the model predictions. The BSCW configuration and geometry make it an ideal test case for generating low-, mid-, and high-fidelity aerodynamic data using various techniques. A schematic of the half-span BSCW is shown in Figure~\ref{fig:aeroelastic_bscw_sketch}.

\subsection{Multi-Fidelity Datasets}

\begin{figure}[!hb]
\centering

\subfigure[Angle of Attack (AoA) and Mach number distribution by fidelity.]{
\includegraphics[trim=0 0 0 0, clip, width=0.65\linewidth]{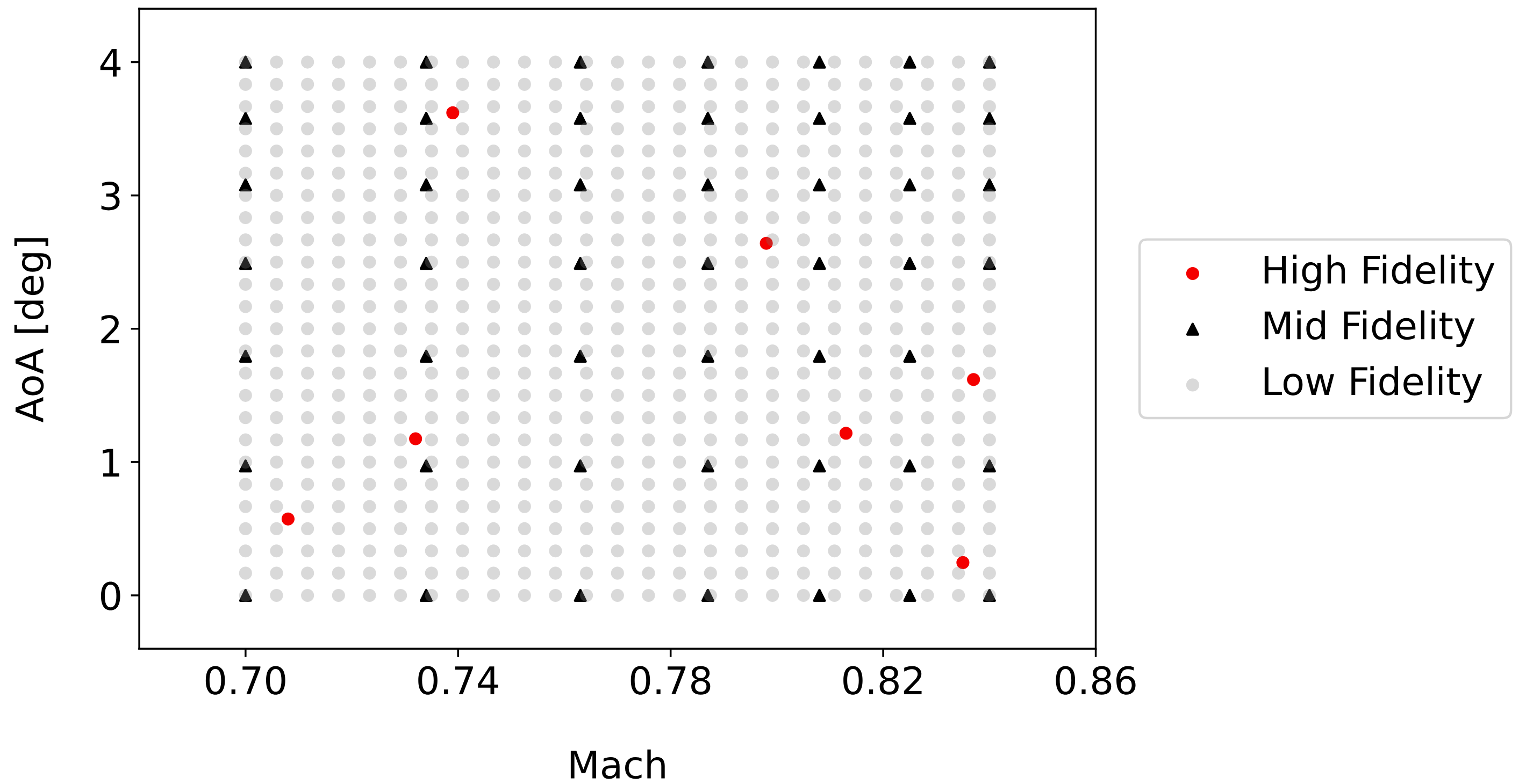}\label{fig:design_space_a}}
\hfill
\subfigure[Predicted aerodynamic coefficients with different fidelities within the design space.]{
\includegraphics[trim=0 0 0 0, clip, width=1\linewidth]{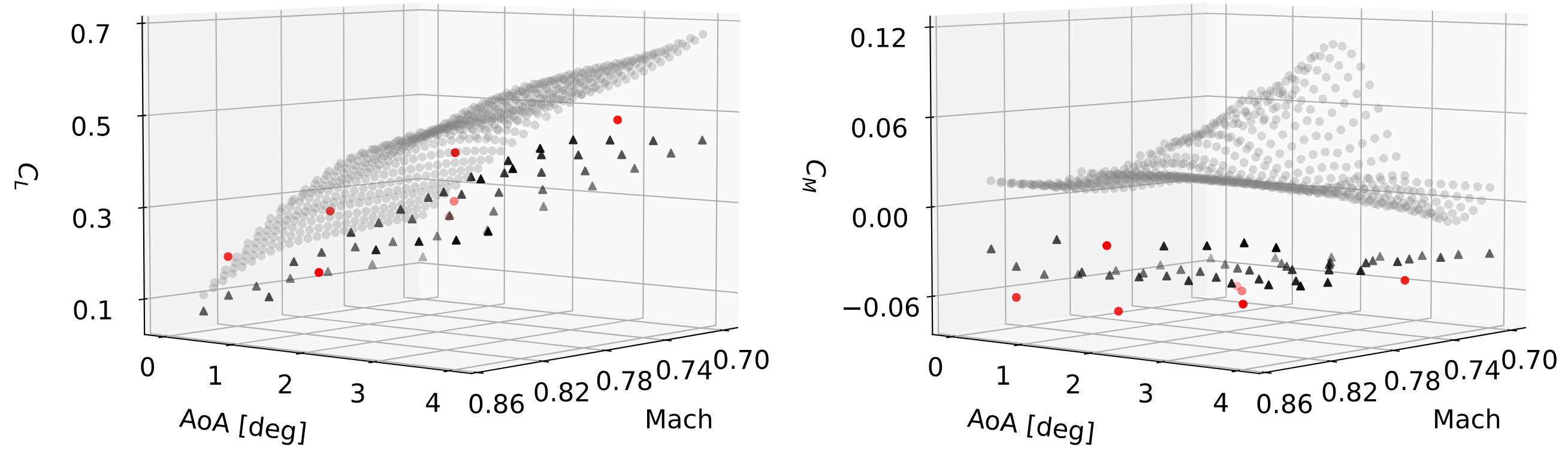}\label{fig:design_space_b}}

\caption{Design variable distribution (AoA-Mach) and aerodynamic coefficient predictions for each fidelity level.}
\label{fig:design_space}
\end{figure}

The datasets include a large number of low-fidelity samples and a limited number of mid- and high-fidelity samples. Low-fidelity data come from panel method, providing quick but approximate solutions. Mid- and high-fidelity data, derived from CFD simulations with different number of grid points, offer detailed and accurate results but are computationally expensive. The combination of these datasets allows for comprehensive training and fine-tuning of the MF-BayNet model. The model uses angle of attack ($AoA$) and Mach number ($M$) as its two independent input parameters. For this study, $AoA$ values ranged from 0 to 4 degrees, and Mach numbers ranged from 0.70 to 0.84 (refer to Figure~\ref{fig:design_space_a}). These specific ranges were selected to capture the critical aerodynamic phenomena occurring in the transonic regime, such as shock wave formation on the wing, and to account for high angles of attack that can lead to boundary-layer separation. This focus ensures that the model is well-suited for analyzing the complex aerodynamic behaviors relevant to the BSCW. Figure~\ref{fig:design_space_b} also highlights the discrepancies in aerodynamic coefficient predicted by each fidelity level, emphasizing the necessity for a model capable of effectively distinguishing and emphasizing the key features of each fidelity.

\subsubsection*{Low-Fidelity}

Low-fidelity data were generated using XFoil, a popular tool for the design and analysis of subsonic airfoils which employs a combination of inviscid panel methods with a boundary layer analysis, allowing it to rapidly generate aerodynamic data. For this study, a NASA SC(2)-0414 airfoil was used. To create a comprehensive dataset, 25 points equally distributed in the design space were used for both $AoA$ and Mach number, resulting in a total of 625 samples. This extensive dataset offers a broad base of quick, approximate aerodynamic solutions. A correction for three-dimensional effects was applied to the XFoil-generated data to enhance its accuracy for the three-dimensional BSCW configuration. These corrections ensure that the low-fidelity data better represent the actual aerodynamic behavior of the wing in three-dimensional flow conditions, making the dataset more valuable for the multi-fidelity model training.

\subsubsection*{Mid-Fidelity}

Mid-fidelity data were obtained from RANS simulations using SU2 v7.5.1 software~\cite{economon2016su2} with a relatively coarse grid of $2.5 \cdot 10^6$ elements. The RANS equations were closed using the one-equation Spalart-Allmaras turbulence model. Convergence was monitored using the Cauchy method applied to the lift coefficient, with a variation threshold of $10^{-7}$ across the last 100 iterations. A $1v$ multigrid scheme was adopted to accelerate convergence. Convective flow discretization utilized the Jameson-Schmidt-Turkel central scheme with artificial dissipation, and flow variable gradients were computed using the Green Gauss method. The biconjugate gradient stabilization linear solver with an ILU preconditioner was selected. The samples were distributed to refine the highly nonlinear regions, particularly at high combinations of $AoA$ and Mach number. This strategic sampling ensures that the mid-fidelity data provide enhanced resolution in critical areas, capturing the complex aerodynamic interactions more effectively. A total of 49 samples were generated, offering more detailed information than low-fidelity data but at a lower computational cost than high-fidelity simulations.

\subsubsection*{High-Fidelity}

The high-fidelity dataset was previously generated~\cite{immordino2024steady}, incurring no further computational cost. It consists of 58 RANS simulations with a fine grid comprising $15.6 \cdot 10^6$ elements. These high-fidelity simulations provide the most detailed and accurate aerodynamic load predictions, serving as a benchmark for validating the MF-BayNet model. For fine-tuning the model, 7 simulations were identified as the minimum number of samples necessary to accurately characterize the discrepancies between mid- and high-fidelity predictions, represented as red dots in Figure~\ref{fig:design_space}. This selection was informed by the authors prior knowledge of the relevant aerodynamic phenomena. The remaining samples were used as a test set to validate the model, demonstrating that with a minimum number of simulations for fine-tuning, the model can achieve a relatively low error on the entire high-fidelity dataset.

\section{Results}\label{sec:results}

The results section presents the performance of the MF-BayNet model. The model predictions are compared to actual high-fidelity results to evaluate accuracy. Metrics such as mean squared error (MSE) and coefficient of determination are used to quantify the performance. Additionally, the MF-BayNet model uncertainty estimates are compared against traditional methods like Co-Kriging, highlighting its superior ability to capture and quantify prediction uncertainty. The results also includes comparisons with BNNs trained on each dataset individually without transfer learning, with model architectures optimized based on the validation set.

\subsection*{Models Performance Comparisons}

Table~\ref{tab_Performance_metrics} compares the performance of MF-BayNet with Co-Kriging and also presents the results of the BNN trained on each dataset individually without transfer learning. The results demonstrate that MF-BayNet significantly outperforms all other models for both $C_L$ and $C_M$. Specifically, MF-BayNet achieves a prediction error ($\epsilon_{cl}$) of 4.2\% and a prediction error ($\epsilon_{cm}$) of 5.5\%, which are approximately half of those obtained with Co-Kriging, which has errors of 7.5\% and 7.7\% for $C_L$ and $C_M$ respectively. Moreover, the standard deviations ($\sigma_{cl}$ and $\sigma_{cm}$) for MF-BayNet are also lower, indicating more consistent and reliable predictions.

In addition, the total prediction error ($\epsilon_{TOT}$) for MF-BayNet is 4.9\%, significantly lower than the 7.6\% observed for Co-Kriging. The BNN models, whether trained on low-fidelity (LF), mid-fidelity (MF), or high-fidelity (HF) datasets, show higher prediction errors and standard deviations, reinforcing the efficacy of the transfer learning and fine-tuning processes utilized by MF-BayNet.

\begin{table}[!ht]
\centering
\small
\begin{tabular}{l c c c c c }
\hline 
\hline 
\textbf{Model Name} & \textbf{$\epsilon_{cl}$\%} & \textbf{$\sigma_{cl}$\%} & \textbf{$\epsilon_{cm}$\%} & \textbf{$\sigma_{cm}$\%} & \textbf{$\epsilon_{TOT}$\%} \\ 
\hline
BNN LF & 14.43 & 1.59 & 42.02 & 1.41 & 31.42 \\
BNN MF & 12.63 & 6.88 & 7.73 & 4.92 & 10.47 \\
BNN HF & 15.29 & 9.00 & 5.62 & 5.96 & 11.52 \\
\hline
CK LF/MF/HF & 7.53 & 2.98 & 7.70 & 9.46 & 7.61 \\
MF-BayNet & \textbf{4.24} & 1.37 & \textbf{5.50} & 0.71 & \textbf{4.91} \\
\hline
\hline 
\end{tabular}
\caption{Performance metrics for different models. Errors ($\epsilon$) and standard deviations ($\sigma$) are calculated on high fidelity test-set.}
\label{tab_Performance_metrics}
\end{table}

\begin{figure}[!htb]
\centering

\subfigure{
\includegraphics[trim=0 0 0 0, clip, width=0.99\linewidth]{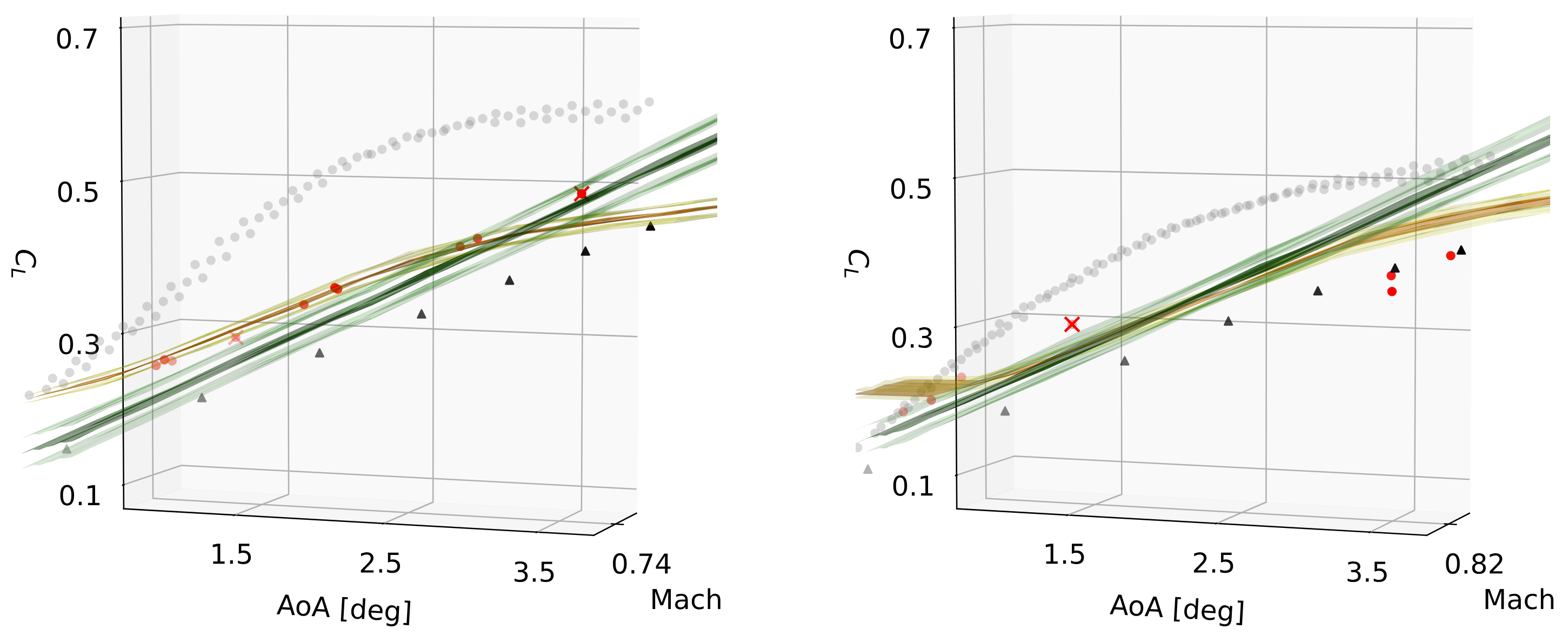}}

\subfigure{
\includegraphics[trim=0 0 0 0, clip, width=0.4\linewidth]{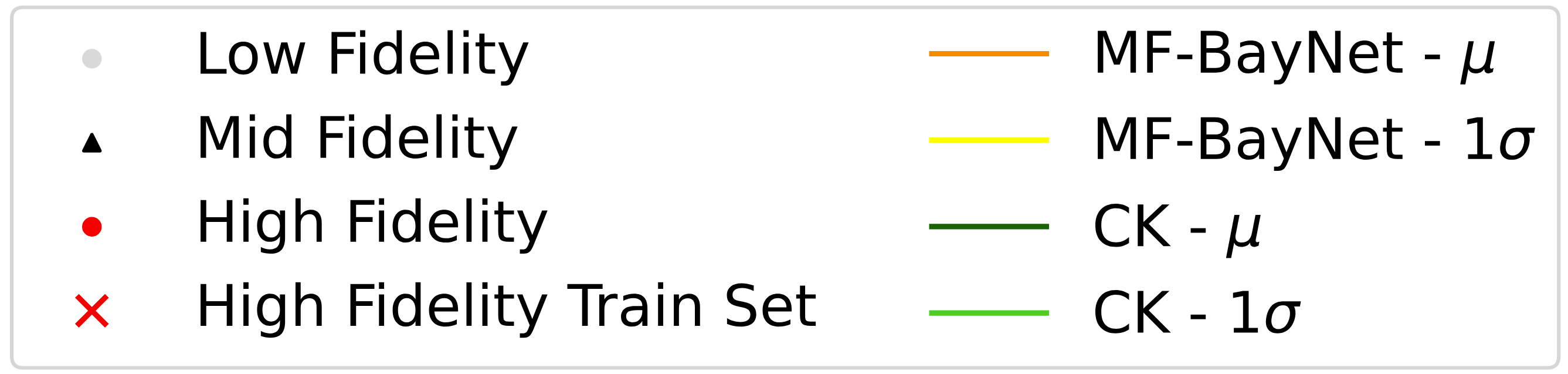}}

\subfigure{
\includegraphics[trim=0 0 0 0, clip, width=0.99\linewidth]{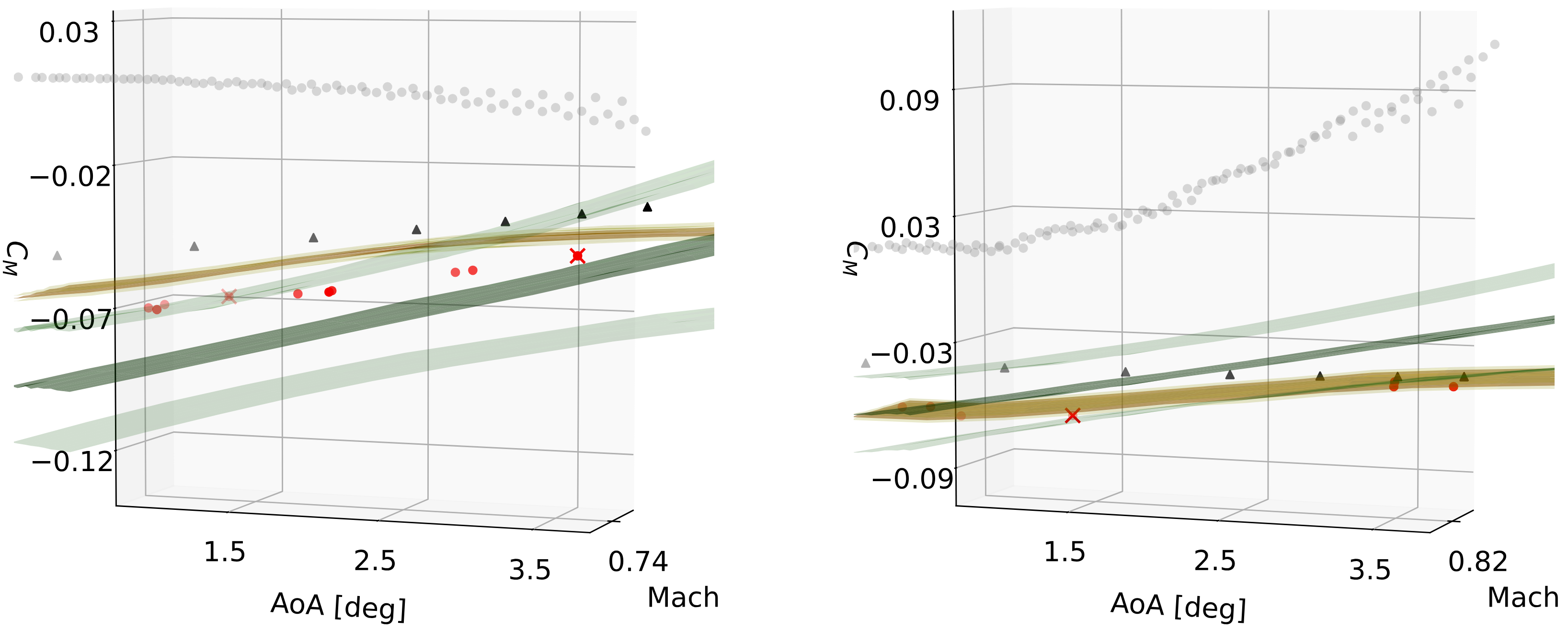}}

\caption{Comparison of MF-BayNet and Co-Kriging predictions for $C_L$ and $C_M$ at two Mach numbers ($M=0.74$ and $M=0.82$) across varying AoA.}
\label{fig:models_comparisons_3d_plots}
\end{figure}

To better illustrate model performance, Figure~\ref{fig:models_comparisons_3d_plots} shows predictions of $C_L$ and $C_M$ at different Mach numbers and AoA. The left panel presents predictions at $M=0.74$, while the right panel shows predictions at $M=0.82$. For $C_L$ predictions, the MF-BayNet model (orange line) closely follows the high-fidelity dataset (red dots), leveraging the knowledge acquired through transfer learning and fine-tuning processes. It exhibits a narrow confidence interval (yellow region) indicating high certainty in its predictions. In contrast, the Co-Kriging (CK) model (drak green line) fails to capture the correct trend, particularly at higher AoA, and exhibits significantly larger confidence intervals (green region). This discrepancy is even more pronounced for $C_M$ predictions. The MF-BayNet model continues to show a good fit with the high-fidelity data, maintaining smaller confidence intervals. However, the Co-Kriging model shows greater uncertainty and higher deviation from the high-fidelity data points, especially noticeable in the larger confidence intervals and trend deviation. This is evident in both $M=0.74$ and $M=0.82$ conditions, highlighting the superior performance and reliability of the MF-BayNet model over Co-Kriging for these aerodynamic predictions.

\subsection*{Uncertainty Quantification Study}

To assess the impact of aleatoric uncertainty on the predictions, we re-trained the Vanilla MF-BayNet model by adding noise to each fidelity dataset separately. The training dataset was augmented by 30\% by adding Gaussian noise with a standard deviation of 1\% of the mean value.

The impact of aleatoric uncertainty on model predictions varies with the fidelity level of the dataset to which noise is added, as reported in Table~\ref{tab_uq}. Generally, introducing noise increases the total error $\epsilon_{TOT}$, except in the case of high fidelity datasets, where $\epsilon_{TOT}$ is lower than the baseline. This can be attributed to the robustness of high fidelity datasets to noise, as they contain more accurate and detailed information, allowing the model to maintain its predictive performance. Additionally, noise in high fidelity data can act as a form of regularization, helping the model generalize better.
Adding noise to medium fidelity datasets results in the most significant increase in $\epsilon_{TOT}$. This higher sensitivity to noise is due to the moderate levels of accuracy and detail in medium fidelity datasets, which are significantly degraded by the introduction of noise. The balance between the inherent detail and the added noise disrupts the model learning process, leading to increased errors.
In contrast, noise in low fidelity datasets produces a mixed impact, either slightly increasing or decreasing $\epsilon_{TOT}$ compared to the baseline. Low fidelity datasets, already less accurate and detailed, show a varied response to noise. In some cases, the noise further degrades data quality, increasing the total error, while in others, it has minimal impact or even aids the model by introducing beneficial variability.
While the standard deviations $\sigma_{cl}$ and $\sigma_{cm}$ exhibit some fluctuation with the addition of noise, these changes are relatively minor compared to the more substantial variations observed in the mean errors $\epsilon_{cl}$ and $\epsilon_{cm}$. This suggests that the model overall uncertainty in its predictions remains relatively stable, even though the accuracy is more sensitive to the noise introduced.

\begin{table}[!htb]
\centering
\small
\begin{tabular}{l c c c c c }
\hline 
\hline 
\textbf{Training Type} & \textbf{$\epsilon_{cl}$\%} & \textbf{$\sigma_{cl}$\%} & \textbf{$\epsilon_{cm}$\%} & \textbf{$\sigma_{cm}$\%} & \textbf{$\epsilon_{TOT}$\%} \\ 
\hline
Vanilla & 4.24 & 1.37 & 5.50 & 0.71 & 4.91 \\

Noisy $C_L$ in LF & 4.33 & 1.30 & 6.19 & 0.72 & 5.34 \\
Noisy $C_L$ in MF & 6.19 & 1.13 & 4.82 & 0.60 & 5.55 \\
Noisy $C_L$ in HF & 3.87 & 1.50 & 5.27 & 0.75 & 4.62 \\

Noisy $C_M$ in LF & 4.38 & 1.03 & 5.04 & 0.56 & 4.72 \\
Noisy $C_M$ in MF & 6.24 & 1.33 & 4.87 & 0.72 & 5.60 \\
Noisy $C_M$ in HF & 3.65 & 1.41 & 4.99 & 0.68 & 4.37 \\

\hline
\hline 
\end{tabular}
\caption{Impact of aleatoric uncertainty on the model predictions. Training dataset augmented by 30\% by adding Gaussian noise with a standard deviation of 1\% of the mean value.}
\label{tab_uq}
\end{table}


\section{Conclusions}\label{sec:concl}

In this study, we have developed a multi-fidelity framework using Bayesian neural networks and transfer learning to enhance the prediction accuracy of aerodynamic loads in a transonic regime over a wing. Our primary objective was to leverage the strengths of BNNs, notably their ability to quantify uncertainty, alongside the efficiency and robustness of TL to improve predictive performance across datasets of different accuracy. The resulting multi-fidelity Bayesian neural network with transfer learning (MF-BayNet) model effectively integrates data of varying fidelities, harnessing the strengths of each fidelity level to produce superior predictive performance and robust uncertainty quantification.

Our results indicate that the MF-BayNet model not only surpasses traditional models trained on single-fidelity datasets but also outperforms state-of-the-art methods such as Co-Kriging. Specifically, MF-BayNet achieves prediction errors that are half of those obtained with Co-Kriging, and exhibits lower standard deviations, indicating more consistent and reliable predictions. These results demonstrate the efficacy of our hybrid approach in achieving superior accuracy and reliability.

Future work will explore the application of this framework to other relevant scenarios, in order to highlight the versatility and robustness of the model. We plan to integrate more advanced techniques for uncertainty quantification, such as Deep Ensembles or Variational Inference, to further improve the reliability and interpretability of the predictions. Also, the incorporation of more diverse and complex datasets will be a key focus, in order to challenge and refine the model performance.


The open-source implementation of the MF-BayNet framework will enable the research community to access, modify, and build upon our work, driving further advancements in the framework and wider applicability.

\bibliographystyle{plain}
\bibliography{biblio.bib}
\end{document}